\documentclass[doublecol]{epl2}% or \documentclass[page-classic]{epl2} for one column style

\usepackage{tabularx}% Table width

\title{Nodeless superconductivity in the SnAs-based van der Waals type superconductor NaSn$_2$As$_2$}
\shorttitle{} %Insert here a short version of the title if it exceeds 70 characters

\author{E. J. Cheng,$^1$ J. M. Ni,$^1$ F. Q. Meng,$^2$ T. P. Ying,$^1$ B. L. Pan,$^1$ Y. Y. Huang,$^1$ Darren Peets,$^1$ Q. H. Zhang,$^2$ and S. Y. Li$^{1,3,*}$\footnote {E-mail: shiyan$\_$li@fudan.edu.cn}}
\shortauthor{E. J. Cheng \etal}

\institute
{$^1$ State Key Laboratory of Surface Physics, Department of Physics, and Laboratory of Advanced Materials, Fudan University, Shanghai 200433, China\\
$^2$ Beijing National Laboratory for Condensed Matter Physics, Institute of Physics, Chinese Academy of Sciences, School of Physical Sciences, University of Chinese Academy of Sciences, Beijing 100190, China\\
 $^3$ Collaborative Innovation Center of Advanced Microstructures, Nanjing 210093, China}


\abstract
{We grew the single crystals of the SnAs-based van der Waals (vdW)-type superconductor NaSn$_2$As$_2$ and systematically measured its resistivity, specific heat, and ultralow-temperature thermal conductivity. The superconducting transition temperature $T_c$ = 1.60 K of our single crystal is 0.3 K higher than that previously reported. A weak but intrinsic anomaly situated at 193 K is observed in both resistivity and specific heat, which likely arises from a charge-density-wave (CDW) instability. Ultralow-temperature thermal conductivity measurements reveal a fully-gapped superconducting state with a negligible residual linear term in zero magnetic field, and the field dependence of $\kappa_0 / T$ further suggests NaSn$_2$As$_2$ is an $s$-wave superconductor.}

\begin{document}

\maketitle

\section{Introduction}

Phonon-mediated conventional superconductors possess $s$-wave pairing symmetry, while unconventional ones prefer $d$-wave, $p$-wave or $s$$_\pm$-wave and their pairing glues are likely not phonon \cite{C0}. Unconventional superconductivity usually resides in quasi-two-dimensional (quasi-2D) compounds, such as cuprates, iron-based superconductors, Sr$_2$RuO$_4$, heavy-fermion and organic superconductors \cite{C1,C2,C3,C4,C5}. In the field of condensed-matter physics, one of the most intriguing themes is finding new quasi-2D superconductors and illuminating their underlying superconducting pairing mechanism.

Quasi-2D superconductors with weak vdW force in interlayers can be exfoliated into monolayer or few atomic-layers films \cite{C6}. By mechanically reducing the dimensionality of a vdW-type superconductor, it is possible to realize a highly crystalline 2D superconducting system with exotic property that differs from the bulk \cite{W1,C7,C8}. For example, Ising superconductivity and a field-induced Bose-metal phase were observed in atomically-thin NbSe$_2$ \cite{C7,C8}. Moreover, with the development of ionic gating techniques, the superconducting transition temperature can be tuned \cite{C9}. To explore more exotic phenomena and clarify the superconducting pairing mechanism in the vdW-type superconductors, different types of compounds are highly desirable.

Recently, the first superconducting 2D SnAs-based compound, NaSn$_2$As$_2$ with a bulk $T_c$ = 1.3 K, was reported \cite{C10}. Later, its sister compound Na$_{1-x}$Sn$_2$P$_2$ with $T_c$ = 2.0 K was also discovered \cite{NaSnP}. As schematically shown in Fig. 1(a), NaSn$_2$As$_2$ consists of SnAs bilayers separated by Na$^+$ cations crystallizing a trigonal $R$$\overline{3}$$m$ unit cell, which is structurally different from the tetragonal ``122" iron-based superconductors \cite{C2}. NaSn$_2$As$_2$ can be exfoliated into monolayer or few-layers films by using liquid-phase or mechanical lift-off techniques, making it a vdW-type superconductor \cite{W1,C11}. Stoichiometric, NaSn$_2$As$_2$ is a non-electron-balanced compound assuming a +2 oxidation state for Sn, and -3 oxidation state for As \cite{C10}. In NaSn$_2$As$_2$, the Sn$^{2+}$ ions are accompanied by lone-pair electrons \cite{C10}. Lone-pair electrons lead to a strong anharmonicity because of the nonlinear terms in the total energy relevant to a large nonhybrid valence electronic contribution, resulting in low thermal conductivity as well as structural distortion \cite{W2}. This is reminiscent of the RE(O,F)BiS$_2$ compounds (RE represents rare earth elements) which includes Bi$^{3+}$ lone pair electrons, and the lone-pair effect is likely relevant to superconductivity \cite{C12}. Recent literature also reported on unconventional superconducting pairing mechanism in RE(O,F)BiS$_2$ compounds \cite{C13}, which contrasts with previous study \cite{W3}. This inspires us to explore the superconducting pairing symmetry of NaSn$_2$As$_2$ with lone-pair electrons.

In this paper, we present the growth of high-quality NaSn$_2$As$_2$ single crystals, and systematic measurements of their resistivity, specific heat, and ultralow-temperature thermal conductivity. The $T_c$ = 1.60 K of our single crystals is 0.3 K higher than that in Ref. \cite{C10}. A previously unobserved anomaly around 193 K is evidenced from both resistivity and specific heat, with its origin may closely related to the formation of CDW. Ultralow-temperature thermal conductivity measurements demonstrate that NaSn$_2$As$_2$ is an $s$-wave superconductor with a nodeless superconducting gap.

\section{Experiments}

Single crystalline NaSn$_2$As$_2$ was synthesized by a Sn-flux method, which is different from that of previous reports \cite{C10,C11}. Arsenic lumps (99.999$\%$, Aladdin), sodium chunks (99.99$\%$, Aladdin), and tin powder (99.999$\%$, Aladdin) in the ratio Na $:$ Sn $:$ As = 1 $:$ 10 $:$ 2 were used as starting materials. A charge of 1.0 g was put in an alumina crucible and sealed in a quartz ampoule. The sealed ampoule was heated to 750 $^{\circ}$C and kept there for 3 days, then cooled at a rate of 1.5 K/h. The ampoule was taken out and decanted with a centrifuge to remove excess Sn flux at 350 $^{\circ}$C, acquiring plate-like single crystals, as shown in the left inset in Fig. 1(c). \\
$\indent$ For transport measurements, the NaSn$_2$As$_2$ single crystal was cleaved and cut into a rectangular shape of dimensions 3.0 $\times$ 0.5 mm$^2$ in the $ab$ plane, with 0.2 mm thickness along the $c$ axis. Four electrodes were directly attached on the sample surface with silver paint, and the typical contact resistance is 30 m$\Omega$ at 0.3 K. Resistivity measurements from 300 to 2 K and down to 0.3 K were performed in a physical property measurement system (PPMS; Quantum Design) and a $^3$He cryostat, respectively. Heat capacity measurements were conducted using the relaxation method in a PPMS equipped with a dilution refrigerator. Microstructure analysis was performed on Tecnai F20 and H-9000NA transmission electron microscopes both equipped with low-temperature holders. The thermal conductivity study was conducted in a dilution refrigerator via a standard four-wire steady-state method with two RuO$_2$ chip thermometers calibrated $in$ $situ$ against a reference RuO$_2$ thermometer. Magnetic fields were normal to the $ab$ plane and the heat current. To obtain a homogeneous field distribution in the NaSn$_2$As$_2$ single crystal, magnetic fields were applied at a temperature above $T_c$ for both resistivity and thermal conductivity measurements.

\section{Results and discussion}

\begin{figure}
\centering
\includegraphics[clip,width=6.2cm]{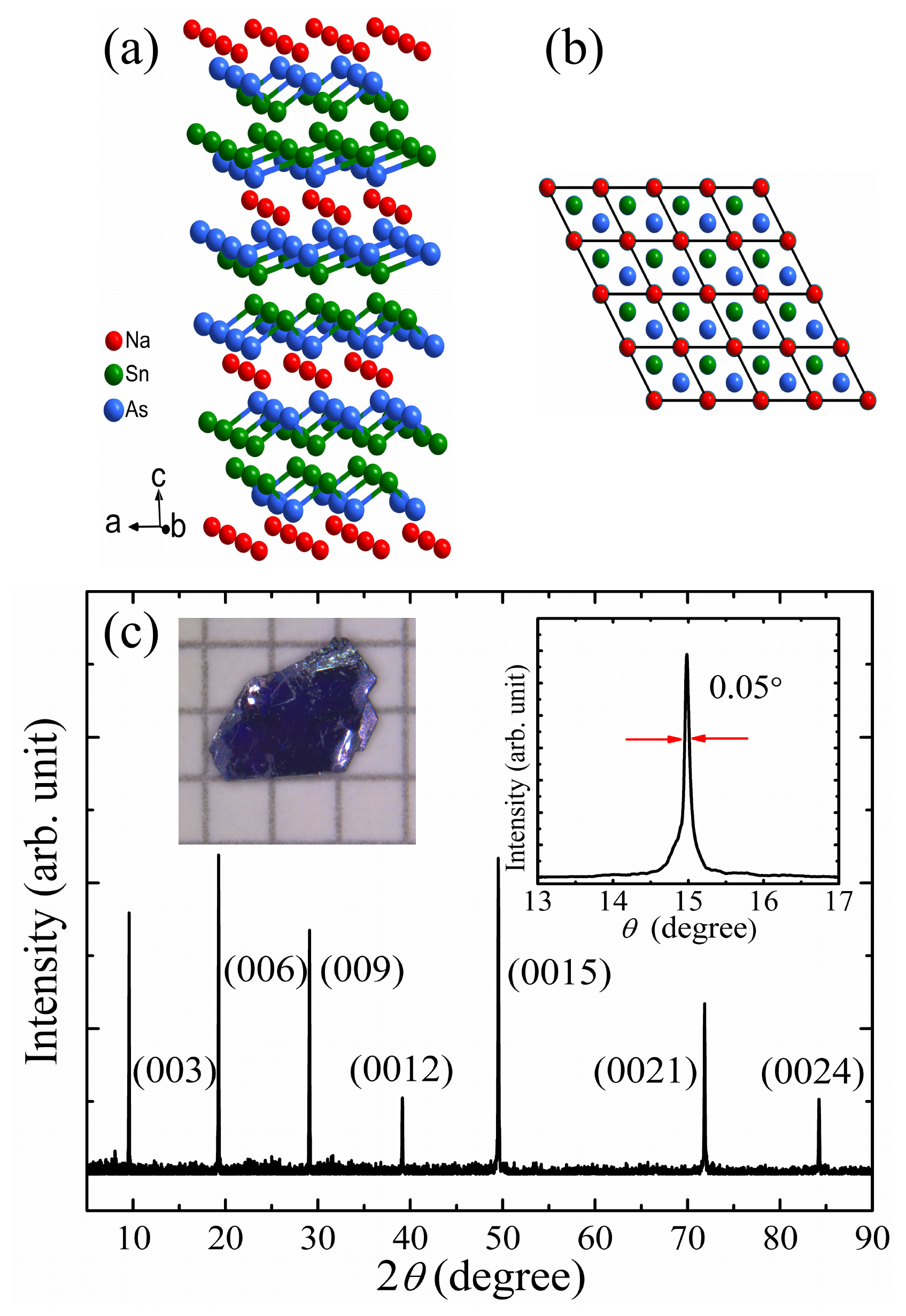}
\caption{(a) Crystal structure of NaSn$_2$As$_2$. Red, green and blue spheres represent Na, Sn and As, respectively. (b) The view of the NaSn$_2$As$_2$ crystal structure down the $c$ axis. (c) X-ray diffraction pattern of the NaSn$_2$As$_2$ single crystal. Left inset: optical image of a typical NaSn$_2$As$_2$ single crystal placed on grid paper with one grid unit 1 $\times$ 1 mm$^2$. Right inset: X-ray rocking curve of (009) Bragg peak having a full width at half maximum of 0.05$^\circ$.}
\end{figure}

From x-ray diffraction measurements, as shown in Fig. 1(c), the largest surface of the single crystals is identified to be the (00$l$) plane of NaSn$_2$As$_2$. From an X-ray rocking curve of the (009) Bragg peak, the full width at half maximum (FWHM) of 0.05$^{\circ}$ indicates the high quality of the NaSn$_2$As$_2$ single crystal. In Fig. 2(a), we present the resistivity of a NaSn$_2$As$_2$ single crystal from 0.3 to 300 K. The resistivity curve from 2 to 10 K is fitted by adopting the formula $\rho$ = $\rho_0$ + $AT^n$, obtaining a residual resistivity $\rho_0$ = 60.3 $\mu$$\Omega$ cm and $n$ = 3.5. The residual resistivity ratio (RRR) $\rho$(300 K)/$\rho_0$ of 3.7 is larger than the 1.6 reported in \cite{C10}, manifesting better sample quality.

In addition to the superconducting transition at low temperature in Fig. 2(a), there are two eye-catching features at the normal state that are worth pointing out and may imply rich physics. The first of all, the fitting factor of the lower region is 3.5 (as shown in the upper inset of Fig. 2(a)), which is higher than the ${T^2}$ of Fermi-liquid behavior but lower than the ${T^5}$ relation caused by phonon scattering. According to the notion proposed by A. H. Wilson, scattering from a low-mass band into a high-density one could induce a higher power law of temperature than that predicted by Fermi-liquid theory in resistivity \cite{C14}. And the plausible scattering mechanism can be charge-density or spin-density fluctuation.
%The same data plotted as $\rho$ $vs$ $T^{3.5}$ is displayed in the upper left of Fig. 2(a). Regardless of the origin of $n$ = 3 for 1$T$-TiSe$_2$ \cite{R1}, the observed 3.5 in NaSn$_2$As$_2$ may be related to some kind of scattering, for example, charge-density fluctuation.

\begin{figure}
\centering
\includegraphics[clip,width=8.6cm]{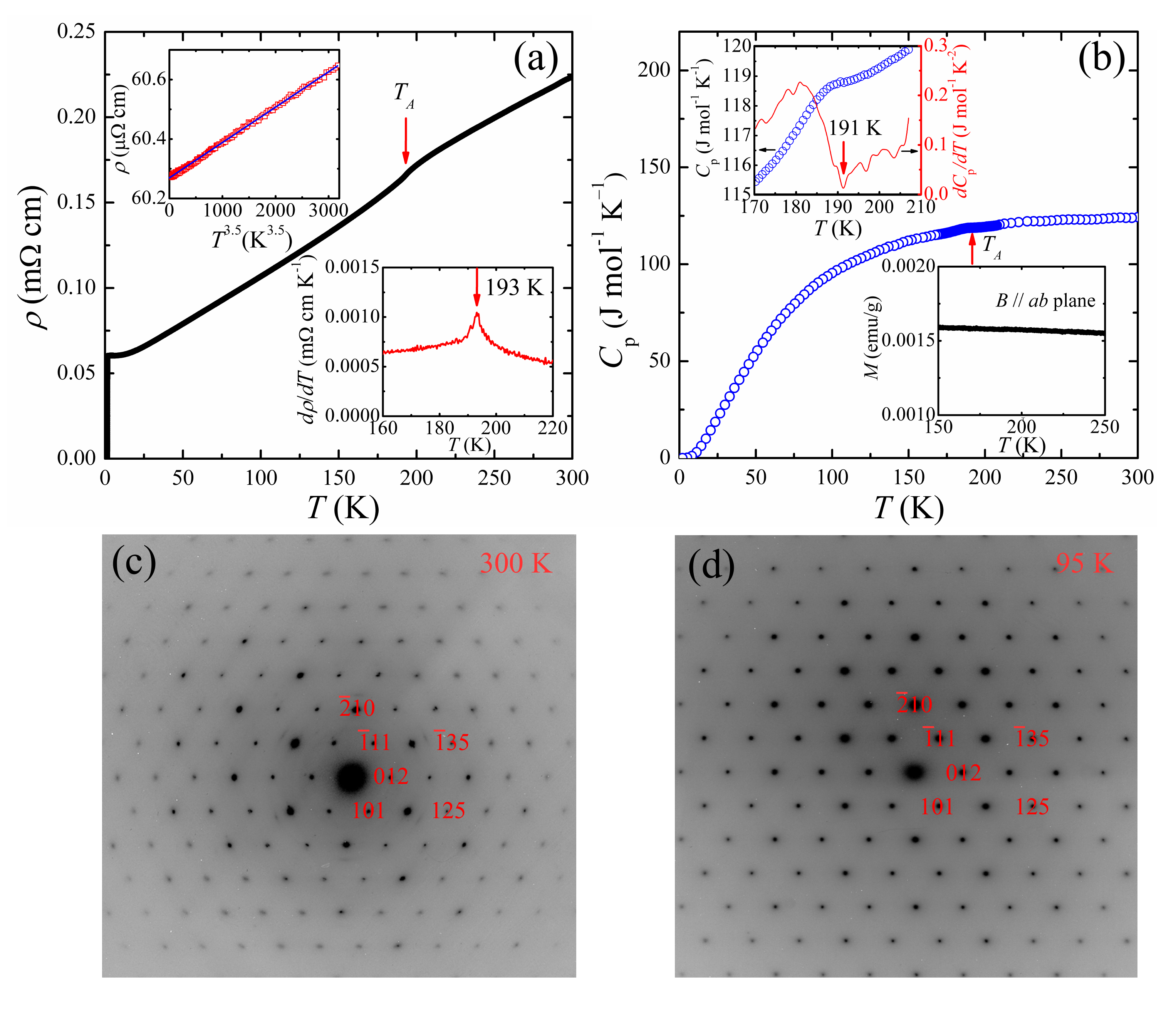}
\caption{(a) Low-temperature resistivity of NaSn$_2$As$_2$ single crystal from 0.3 to 300 K; the weak anomaly at 193 K is marked as $T_{A}$. The inset in the upper left demonstrates the low-temperature power law by ploting $\rho$ $vs$ $T^n$, with $n$ = 3.5. Right inset shows the derivative of resistivity with respect to temperature together across the kink at 193 K. (b) Specific heat of a NaSn$_2$As$_2$ single crystal. A weak phase transition at 191 K is shown and marked with a red arrow. The left inset shows the phase transition and the derivative of the specific heat with respect to temperature. The right inset displays the susceptibility result with a magnetic field of 0.5 T. (c) and (d) show the electron diffraction patterns of NaSn$_2$As$_2$ at 300 and 95 K, respectively.}
\end{figure}

\begin{figure}
\centering
\includegraphics[clip,width=8.750cm]{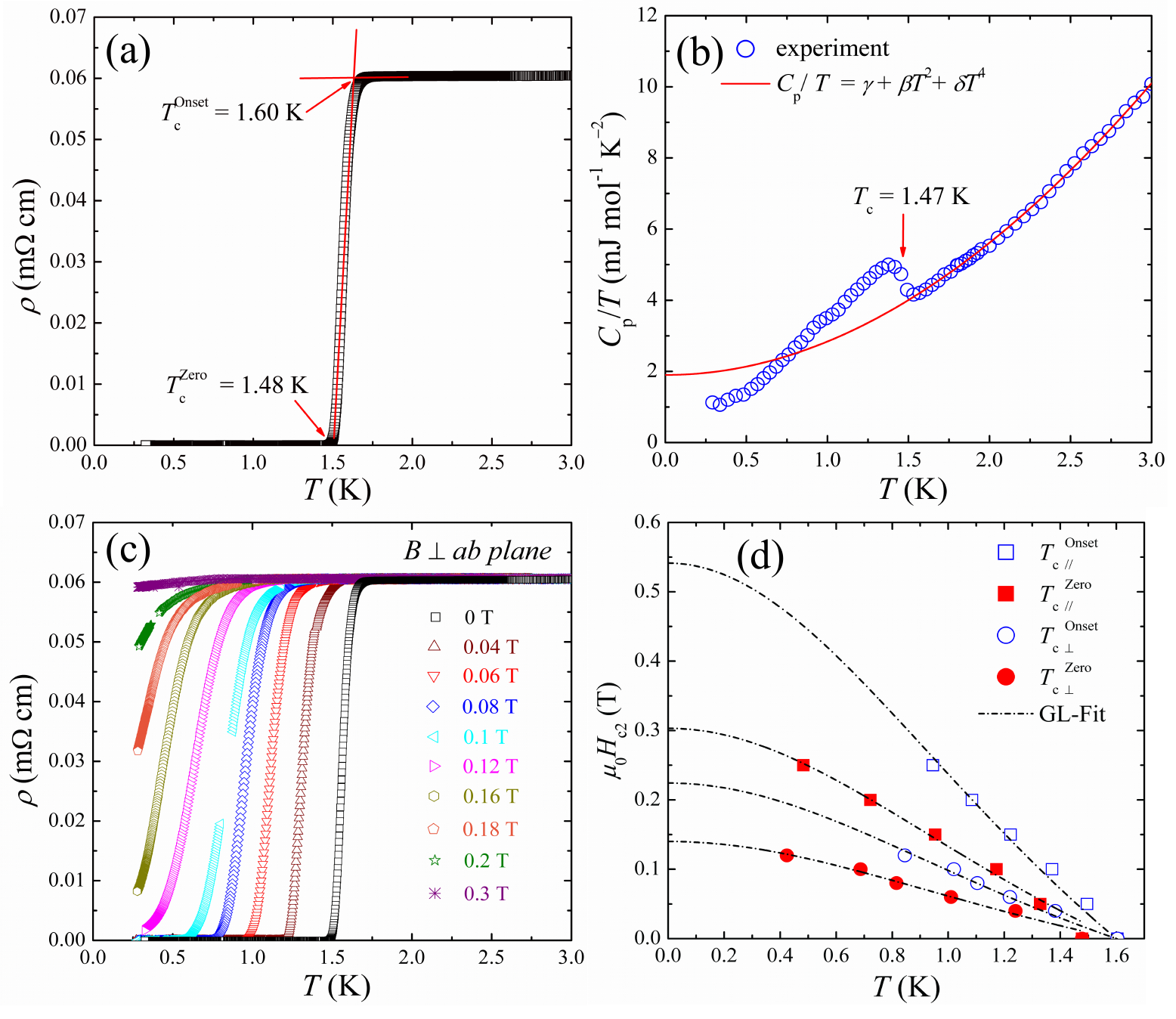}
\caption{(a) Low-temperature resistivity of NaSn$_2$As$_2$ single crystal in zero field. The $T_c$ is defined as the intersection of two straight lines, which gives 1.60 K and 1.48 K for onset and zero-point $T_c$s, respectively. (b) Temperature dependence of specific heat divided by temperature at low temperature. The red line shows the fitting above the superconducting state with  $C_p$/$T$ = $\gamma$ + $\beta$$T^2$ + $\delta$$T^4$. (c) Low-temperature resistivity at magnetic fields perpendicular to the $ab$ plane. (d) The upper critical fields defined by onset and zero-point superconducting temperatures with field perpendicular and parallel to the $ab$ plane.}
\end{figure}

The second prominent feature is a kink situated at 193 K, which has not been reported before \cite{C10}. This anomaly is in consistent with our specific heat experiment where a weak peak around 191 K is observed, as shown in Fig. 2 (b). Notwithstanding the weak signal, we incline to think this kink is intrinsic, if considering the high quality of our single crystals and the bulk sensitive of the specific heat measurement. Contrast to the sharp {``$\lambda$ "}-shape peak associated with structural phase transition, this broad peak should not originate from a first-order structural phase transition. In order to further verify that the anomaly is not rooted in a structural phase transition, we carried out the room- and low-temperature electron diffraction experiments. Figure 2(c) displays the diffraction pattern of NaSn$_2$As$_2$ at 300 K, which can be indexed to a hexagonal unit cell which agrees well with the crystalline structure of NaSn$_2$As$_2$. Typical diffraction spots, such as (101), (012), ($\overline{1}$11), ($\overline{2}$10), ($\overline{1}$35), and (125), can be well indexed. Below the anomaly temperature, the diffraction pattern of 95 K is almost identical to that of 300 K, excluding a structural phase transition.

To check if the anomaly arises from a magnetic phase transition, the susceptibility measurement was conducted. As shown in the inset of Fig. 2(b), no anomaly around 193 K appears. Recent electronic structure calculations on NaSn$_2$As$_2$ found that its electronic states near fermi level ($E_F$) are primarily derived from the contribution of nonmagnetic Sn-$s$, Sn-$p$, and some As-$p$ orbitals \cite{C11}, which further suggests that this compound lies far from magnetic instabilities. CDWs often appear in 2D compounds \cite{C15}, and are accompanied by superlattice \cite{C15,C16}. Together with a weak broad peak in specific heat and a higher power law of temperature in resistivity, we speculate that the anomaly in 2D NaSn$_2$As$_2$ might arise from a CDW instability. The failed observation of a superlattice pattern at low temperature should be due to the weakness of the CDW in NaSn$_2$As$_2$. To verify this speculation, more solid evidence will be required.

The superconducting transition in the resistivity is plotted in Fig. 3(a), and the $T_c$s are 1.60 and 1.48 K for the onset and zero-point ($T_c^{Onset}$, $T_c^{Zero}$) temperatures, respectively. Figure 3(b) shows the low-temperature specific heat divided by the temperature, $C_p/T$, as a function of temperature in zero field. The superconducting transition at 1.47 K corresponds to the observation in resistivity. Above $T_c$, the data from 1.5 to 3 K of $C_p$/$T$ $vs$ $T$ can be well fitted by $C_p$/$T$ = $\gamma$ + $\beta$$T^2$ + $\delta$$T^4$. The electronic specific heat coefficient $\gamma$ and the phononic coefficient $\beta$ are determined to be 1.90 mJ mol$^{-1}$ K$^{-2}$ and 0.94 mJ mol$^{-1}$ K$^{-4}$, respectively. The Debye temperature $\Theta_D$ $ \approx $ 218 K is estimated by adopting the formula $\Theta_D$ = (12$\pi^4$$r$$N_A$$k_B$/5$\beta$)$^{1/3}$, where $r$ = 5 is the number of atoms per formula unit, $N_A$ is the Avogadro constant, and $k_B$ is the Boltzmann constant, respectively. Figure 3(c) plots the low-temperature resistivity of NaSn$_2$As$_2$ in various magnetic fields up to 0.3 T showing that the superconducting transition is gradually suppressed with magnetic field. Considering the Ginzburg-Landau theory, $\mu_0$$H_{c2}$($T$) = $\mu_0$$H_{c2}$(0) (1-{$t^2$})/(1+{$t^2$}), here $t$ = $T$/$T_c$, the zero-temperature upper critical field $\mu_0$$H_{c2}$(0) is estimated. As shown in Fig. 3(d), the upper critical fields $H_{c2}$s defined by the onset and zero-point temperatures ($T_{c\bot}^{Onset}$ and $T_{c\perp}^{Zero}$) are 0.22 and 0.14 T, respectively, for magnetic fields normal to the $ab$ plane. The $H_{c2}$s with field parallel to the $ab$ plane are also yielded, and the values give 0.54 and 0.30 T for onset and zero-point temperatures, respectively. Based on these values of $H_{c2}$s, the superconducting state of quasi-2D NaSn$_2$As$_2$ is anisotropic.

\begin{figure}
\centering
\includegraphics[clip,width=6cm]{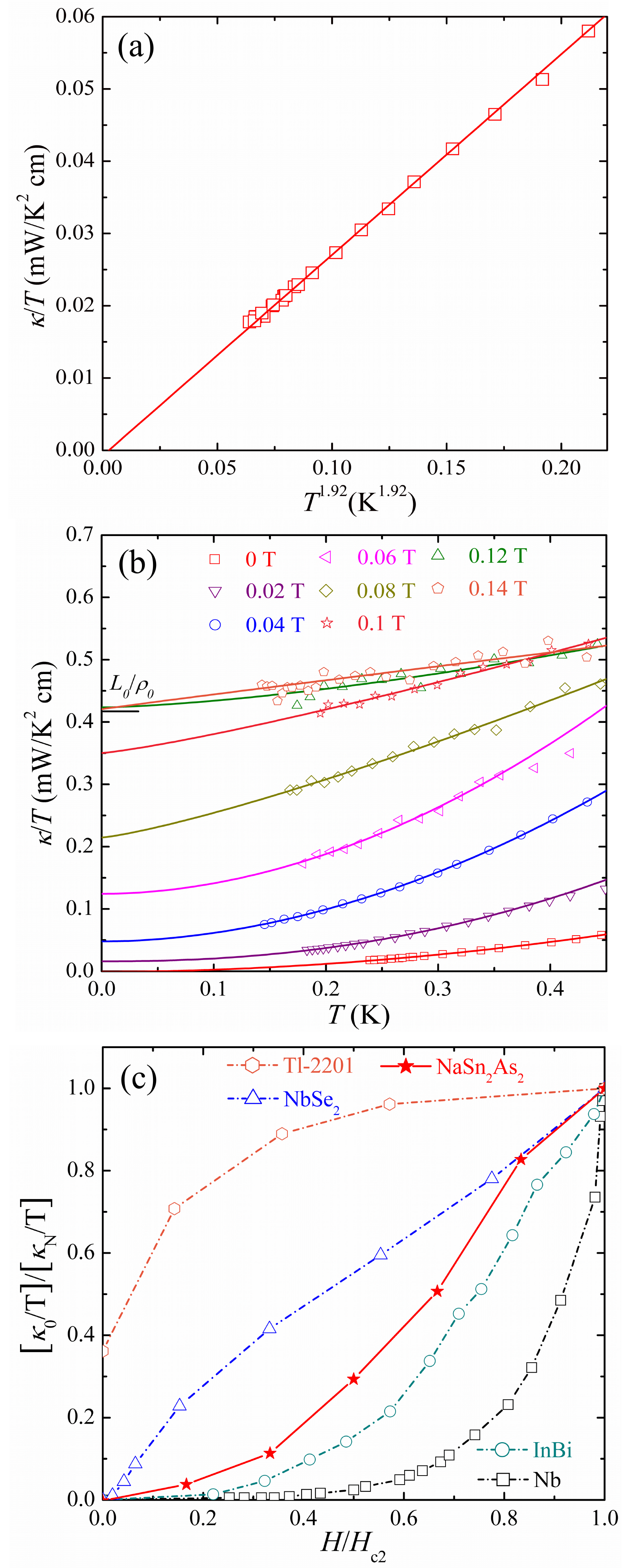}
\caption{(a) Low-temperature thermal conductivity of NaSn$_2$As$_2$ single crystal in zero field and (b) magnetic fields. The solid lines in both zero field and fields represent the fitting according to the formula of $\kappa/T$ = $a$ + $bT^{\alpha-1}$. In zero field, the value of the residual linear term $\kappa_0/T$ is - 1 $\pm$ 1 $\mu$W / K$^2$ cm, which is negligible. In an applied  field at 0.12 T, $\kappa_0/T$ meets the normal-state Wiedemann-Franz law expectation $L_0$/$\rho_0$. (c) Normalized residual linear term $\kappa_0/T$ of NaSn$_2$As$_2$ as a function of $H$/$H_{c2}$, with bulk $H_{c2}$ = 0.12 T. Data on the clean $s$-wave superconductor Nb \cite{C18}, the dirty $s$-wave superconductor InBi \cite{C19}, the multiband $s$-wave superconductor NbSe$_2$ \cite{C20}, and the overdoped $d$-wave cuprate superconductor Tl-2201\cite{C21} are included for comparison.}
\end{figure}

Ultralow-temperature thermal conductivity measurement is an established bulk technique to probe the superconducting gap structure \cite{C17}. In unconventional superconductors, symmetry-imposed nodes are often observed within the superconducting gap \cite{C17}. In our experiment, thermal conductivity can be separated into two contributions, $\kappa_e$ and $\kappa_p$, associated with the one from electrons and phonons, respectively. In order to study their specific contributions, the formula $\kappa/T$ = $a$ + $bT^{\alpha-1}$ \cite{C17} is adopted for fitting, with the two terms $aT$ and $bT^{\alpha}$ represent contributions from electrons and phonons, respectively. The residual linear term $\kappa_0/T$ of - 1 $\pm$ 1 $\mu$W K$^{-2}$ cm $^{-1}$ is obtained by extrapolating the $\kappa/T$ to zero temperature (See Fig. 4(a)). Noting that a negative $\kappa_0/T$ intercept is unphysical, the value of $\kappa_0/T$ is negligible with our experimental error bar of $\pm$ 5 $\mu$W K$^{-2} $cm$^{-1}$.

As for the parameter of $\alpha$, its value is typically between 2 and 3 because of specular reflections of phonons at the sample surfaces \cite{C17}. For $s$-wave nodeless superconductors, there are no fermionic quasiparticles to conduct heat as $T$ $\rightarrow$ 0, since all electrons become Cooper pairs \cite{C17}. Therefore, there is no residual linear term of $\kappa_0/T$, as seen in Nb, InBi and NbSe$_2$ \cite{C18,C19,C20}. However, for nodal superconductors, a substantial $\kappa_0/T$ in zero field contributed by the nodal quasiparticles has been found. For example, $\kappa_0/T$ of the overdoped ($T_c$ = 15 K) $d$-wave cuprate superconductor Tl$_2$Ba$_2$CuO$_{6+\delta}$(Tl-2201) is 1.41 mW K$^{-2}$ cm$^{-1}$, $\sim$ 36$\%$ $\kappa_{N0}/T$  \cite{C21}. For the $p$-wave superconductor Sr$_2$RuO$_4$ ($T_c$ = 1.5 K), $\kappa_0/T$ = 17 mW K$^{-2}$ cm$^{-1}$ was reported, more than 9$\%$ $\kappa_{N0}/T$  \cite{C3}. Hence, the negligible $\kappa_0/T$ of NaSn$_2$As$_2$ strongly suggests a nodeless superconducting gap structure.

Figure 4(b) shows $\kappa/T$ $vs$ $T$ plots of temperature-dependent thermal conductivity for a NaSn$_2$As$_2$ single crystal under magnetic fields. All the data of $\kappa/T$ $vs$ $T$ are fitted and the $\kappa_0/T$ for each field is obtained. In 0.12 and 0.14 T, the value of $\kappa_0/T$ is 0.43 $\pm$ 0.04 and 0.42 $\pm$ 0.04 mW K$^{-2}$ cm$^{-1}$, respectively. We determined the normal-state expectation value of the Wiedemann-Franz law $L_0$/$\rho_0$ (0.12 T) = 0.41 mW K$^{-2}$ cm$^{-1}$, where the Lorenz number $L_0$ = 2.45 $\times$ 10$^{-8}$ W $\Omega$ K$^{-2}$ and $\rho_0$ (0.12 T) = 60.3 $\mu$$\Omega$ cm. The value of $\kappa_0/T$ in 0.12 T meets the expectation, which means that the normal state has been reached. Note that the $H_{c2}$ defined by $\rho$ = 0 is 0.14 T, when magnetic field perpendicular to the $ab$ plane. Therefore, the upper critical fields of NaSn$_2$As$_2$ in our thermal conductivity and resistivity experiments are consistent, which demonstrates that our thermal conductivity measurements are reliable.

Further information on the superconducting pairing symmetry can be provided by examining the behavior of field-dependent $\kappa_0(H)/T$ as a function of $H$/$H_{c2}$ \cite{C22}, as shown in Fig. 4(c). Data on the clean $s$-wave superconductor Nb \cite{C18}, the dirty $s$-wave superconductor InBi \cite{C19}, the multiband $s$-wave superconductor NbSe$_2$ \cite{C20}, and the overdoped $d$-wave cuprate superconductor Tl-2201 \cite{C21} are plotted for comparison. For the single-band clean $s$-wave superconductor Nb, $\kappa_0(H)/T$ grows exponentially with field \cite{C18}, while for the $s$-wave InBi in the dirty limit, the curve is exponential at low $H$ and displays roughly linear behavior closer to $H_{c2}$ \cite{C19}. For nodal superconductor Tl-2201, a small field can yield a quick growth due to the Volovik effect, and the low-field $\kappa_0(H)/T$ shows a roughly $\sqrt{H}$ dependence \cite{C21}. In the case of NbSe$_2$, the distinct $\kappa_0(H)/T$ behavior was well explained by multiple superconducting gaps with different magnitudes \cite{C20}.

By comparing the curve of the normalized $\kappa_0(H)/T$ for NaSn$_2$As$_2$ with others, the field dependence of $\kappa_0(H)/T$ most closely resembles that of InBi. The faster growth has two possible explanations. One is that NaSn$_2$As$_2$ is a multiple-band superconductor. However, from ARPES experiments, there is only a hole-type band crossing $E_F$, and hole carriers dominate the density of states of the Fermi surface at low temperature \cite{C11}. Therefore, we exclude this explanation. The other possibility is that NaSn$_2$As$_2$ is a ``dirty" superconductor. However, due to the lack of fermi velocity in NaSn$_2$As$_2$, we can not derive the electron mean free path $l$. Thus, it is an open question whether NaSn$_2$As$_2$ is an $s$-wave superconductor in the dirty limit.

\section{Summary}

In summary, we synthesized NaSn$_2$As$_2$ single crystals using a Sn-flux method, and a superconducting transition temperature $T_c$ = 1.60 K is found. From ultralow-temperature thermal conductivity experiments, a fully-gapped superconducting state has been revealed. The field dependence of $\kappa_0 / T$  further confirms that NaSn$_2$As$_2$ is an $s$-wave superconductor. A weak anomaly at 193 K is discovered, and we propose that the anomaly might arise from a CDW instability. Further work is needed to determine whether the anomaly results from a CDW instability, and illuminate the interplay between CDW and superconductivity with lone-pair electrons.\\

\acknowledgments

This work is supported by the Ministry of Science and Technology of China (Grant No. 2015CB921401 and 2016YFA0300503), the Natural Science Foundation of China (Grant No. 11422429 and 11421404), and the NSAF (Grant No. U1630248).\\

\end{document}